\documentclass[aps,prd,superscriptaddress,showkeys,preprint,nofootinbib]{revtex4}
\usepackage{graphics,epsfig}
\usepackage{subfigure}
\usepackage{float}
\usepackage{epstopdf}
\usepackage{graphicx}
\usepackage{dcolumn}
\usepackage{amsmath}
\usepackage{epstopdf}
\usepackage{svg}
\usepackage[colorlinks=true, linkcolor=blue, citecolor=blue, urlcolor=blue]{hyperref}
\newcommand{\orcid}[1]{\href{https://orcid.org/#1}{\includegraphics[width=8pt]{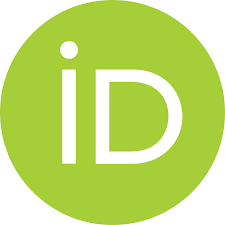}}}
\usepackage{nameref}
\usepackage{tikz}

\begin{document}
\title{Propagation of massless particles around a BTZ-ModMax\\
black hole}

\author{Shubham Kala\,\orcid{0000-0003-2379-0204}}
\email{shubhamkala871@gmail.com}
\affiliation{The Institute of Mathematical Sciences, C.I.T Campus, Taramani-600113, Chennai, Tamil Nadu, India }

\begin{abstract}
We investigate the deflection of light around a BTZ-ModMax black hole, focusing on the influence of the ModMax parameter and the cosmological constant. The trajectories of massless particles are explored through an analysis of null geodesics, providing deeper insights into the gravitational lensing effects in this modified black hole geometry. Using the Gauss-Bonnet theorem, we derive the deflection angle of light and provide a detailed examination of how the ModMax parameter and the cosmological constant impact the bending of light. Furthermore, we compare the deflection results obtained for the BTZ-ModMax black hole with those of the charged and static BTZ black hole solution, highlighting key differences and insights into the role of nonlinear electrodynamics in the gravitational lensing phenomena. This study provides valuable insights into the deflection of light in modified black hole solutions within the framework of lower-dimensional geometry, highlighting the complex relationship between nonlinear dynamics and spacetime curvature.
\end{abstract}


\maketitle

\section{Introduction}
\label{introduction}
Nonlinear electrodynamics (NLED) theories are generalizations of Maxwell's theory that address phenomena beyond its scope, such as the self-interaction of virtual electron-positron pairs and the elimination of spacetime singularities in black holes (BHs) \cite{Heisenberg:1936nmg,schwinger1951gauge,Yajima:2000kw}. These theories also influence gravitational redshift around strongly magnetized compact objects and eliminate the cosmological singularity associated with the Big Bang \cite{Baccigalupi:2002bh,Ayon-Beato:1999qin,DeLorenci:2002mi,Dymnikova:2004zc,Corda:2009xd}. Examples of NLED theories include Born-Infeld (BI) \cite{Born:1934gh}, Euler-Heisenberg (EH) \cite{Heisenberg:1936nmg}, and Power-Law (PL) \cite{Hassaine:2007py,Maeda:2008ha}. Each theory has distinct properties, such as electromagnetic duality, conformal invariance, and the resolution of self-energy divergences in point particles. Recently, the Modified-Maxwell (ModMax) NLED theory was introduced \cite{bandos2020nonlinear}, distinguished by the dimensionless parameter \(\gamma\), which recovers Maxwell's theory for \(\gamma = 0\) \cite{kosyakov2020nonlinear}. Coupling ModMax electrodynamics with gravity has led to various BH solutions, expanding our understanding of spacetime and field interactions \cite{Kruglov:2021bhs,Kuzenko:2021qcx,Avetisyan:2021heg}.
The first-ever BH solution in (2+1)-dimensional gravity within the framework of General Relativity (GR) was proposed by Ba$\tilde{n}$ados, Teitelboim, and Zanelli in 1992, and it is widely recognized as the BTZ BH \cite{Banados:1992wn}. This BH solution features a negative cosmological constant and exists in lower dimensions, it has opened a new and exciting field of study in theoretical physics, offering unique insights into lower dimensional gravity. The BTZ BH, unlike (3+1)-dimensional solutions such as Schwarzschild and Kerr, is asymptotically anti-de Sitter and has no curvature singularity at the origin \cite{Banados:1992gq}. However, it retains key features like an event horizon, Hawking temperature, and well-defined thermodynamic properties, offering valuable insights into lower-dimensional gravity. \cite{Quevedo:2008ry}. Then, the study of gravity in three-dimensional spacetime absorbed a lot of attentions due to different aspects of their physics properties.  In this regard, different three-dimensional BHs have been obtained in Einstein’s gravity and also modified theories of gravity which are coupled with linear and nonlinear matters \cite{Cardenas:2014kaa,HosseinHendi:2017soi,Nashed:2017fnd,Mu:2019jjw,Canate:2020btq,EslamPanah:2022ihg,Karakasis:2023ljt,EslamPanah:2023rqw,Barrientos:2022bzm,Barrientos:2024umq,Arenas-Henriquez:2022www,Arenas-Henriquez:2023hur,Diaz:2024wbh}. More recentaly, an analytic BTZ BH solution is obtained by coupling Einstein’s gravity with the ModMax NLED field followed by detailed thermal properties \cite{eslam2024thermodynamics}.  
Geodesics are fundamental to the study of spacetime geometry, as they describe the trajectories of test particles and light rays in a curved spacetime \cite{Misner:1973prb}. They provide critical information about the causal structure of spacetime, the nature of gravitational interactions, and the response of the spacetime to the presence of mass and energy, serving as a key tool for understanding the dynamics of relativistic systems. In the context of a (2 + 1)-dimensional BH, geodesics describe the trajectories followed by massless and massive particles under the influence of the BH's gravitational field \cite{Cruz:1994ir}. The detailed study of geodesics structure provide essential insights into the behavior of light and matter around the BH, revealing properties like the deflection of light, event horizon structure, and thermodynamic characteristics. Understanding geodesics in such lower-dimensional spacetimes helps to explore fundamental aspects of gravity and spacetime curvature in reduced dimensions. The analysis of geodesics in lower-dimensional frameworks, as explored by previous authors, is referenced herein \cite{Dasgupta:2012zf,Rahaman:2013gw,Xu:2014uka,Soroushfar:2015dfz,Hendi:2020yah,Kala:2021ppi,Battista:2022krl}.
Over the past few decades, gravitational lensing (GL) has been extensively studied due to its profound implications in astrophysics. The gravitational deflection of light was first quantitatively analyzed by Soldner in 1801 using Newtonian mechanics \cite{jaki1978johann}. In 1959, Darwin extended this analysis and obtained the deflection angle of light in strong-field regime using the Schwarzschild metric \cite{darwin1959gravity}. Further studies explored the deflection angle and the formation of images due to the Schwarzschild BH, with expressions involving elliptic integrals of the first kind. Virbhadra and Ellis introduced a lens equation for strong GL by a Schwarzschild BH and proposed a method to compute the bending angle \cite{Virbhadra:1999nm}. Iyer et al., formulated an analytical perturbative approach to calculate the bending angle of light rays, lensed by a Schwarzschild BH \cite{Iyer:2006cn}. W. Rindler and M. Ishak \cite{Rindler:2007zz} investigated the influence of the cosmological constant on the deflection of light, providing critical insights into its role in gravitational lensing phenomena. Gibbons and Werner introduced a straightforward yet elegant method for studying GL, utilizing the Gauss-Bonnet theorem (GBT) to derive the deflection angle through the Gaussian optical curvature of spherically symmetric spacetime \cite{Gibbons:2008rj}. Bozza further advanced the understanding of strong lensing by considering a spherically symmetric BH, where an infinite series of higher-order images are formed, later extending the framework to include rotating BHs \cite{Bozza:2010xqn}. Furthermore, Ishihara et al. \cite{Ishihara:2016vdc}, extended the Gibbons-Werner method to spacetimes with rotating BHs, using the optical geometry approach. Their method provided a more intuitive framework for understanding the deflection of light in curved spacetime, highlighting the connection between geometry and the bending of light in gravitational fields. Ono et al. \cite{Ono:2019hkw}, studied the effects of finite distance on the gravitational deflection angle of light. Fathi et al. \cite{Fathi:2019jid}, investigated the deflection angle of massless particles in a static BTZ BH background within the framework of scale-dependent gravity. Additionally, Kala et al.~\cite{Kala:2020viz}, derived an exact expression for the bending angle of light around a rotating BTZ BH. Furthermore, Upadhyay et al. \cite{Upadhyay:2023yhk}, analyzed the weak deflection angle of a charged massive BTZ BH. Notably, the study of light propagation in the vicinity of ModMax BH in $(3+1)$-dimensional geometry  reported by Herrera et.al \cite{Guzman-Herrera:2023zsv}. Numerous other studies have also explored the deflection of massless particles in lower as well as $(3+1)$-dimensional spacetimes, with relevant references provided herein \cite{Nakashi:2019jjj,Gonzalez:2019xfr,Narzilloev:2021jtg,Sultan:2024zmp,Jusufi:2017vta,Sakalli:2017ewb,Ovgun:2018fte,Ovgun:2018tua,Javed:2019kon,Javed:2020lsg,Javed:2020wsv,Kala:2020prt,Kala:2022uog,Javed:2022rrs,Pantig:2024ixc,Vishvakarma:2024icz,Kala:2024fvg,Turakhonov:2024smp,Kala:2025xnb,Roy:2025hdw,Roy:2025qmx,Kukreti:2025rzn,Kala:2025fld,Ahmed:2025ylr,Pandey:2025wnk}. Recently, several studies have explored the thermodynamic and optical properties of ModMax BHs in $(3+1)$-dimentional spacetime, highlighting their rich structure and physical implications~\cite{EslamPanah:2024tex,EslamPanah:2024fls,EslamPanah:2025bfh,NooriGashti:2025rwl,panah2025some}. These works emphasize how the ModMax parameter influences BH geometry, stability, and observable quantities, thereby motivating the present analysis in a $(2+1)$-dimensional setting. In this paper, we aim to investigate various aspects of light propagation around a ModMax BH, including null geodesics, possible trajectories, the stability of null geodesics, Gaussian optical curvature, and the deflection angle of light. The outline of this manuscript is as follows: In Section \ref{sec2}, we examine the BTZ-ModMax BH spacetime, focusing on its horizon structure in Subsection \ref{sec2.1}. Section \ref{sec3} discusses null geodesics, including their stability (Subsection \ref{sec3.1}) and the Gaussian optical curvature (Subsection \ref{sec3.2}). In Section \ref{sec4}, we investigate the deflection of light around this spacetime. Finally, Section \ref{sec5} summarizes the key findings and conclusions of this work.
\section{BTZ-ModMax BH Spacetime}
\label{sec2}
\subsection{Field equation for ModMax NLED theory}
The action describing Einstein's theory of gravitation coupled with ModMax NLED and the cosmological constant in a three-dimensional spacetime is given by \cite{bandos2020nonlinear},
\begin{equation}
I = \frac{1}{16\pi} \int_{\partial M} d^3x \, \sqrt{-g} \left[ R - 2\Lambda - 4 \mathcal{L} \right],
\end{equation}
where \( R \) is the Ricci scalar, and \( \Lambda \) is the cosmological constant. The parameter \( g = \det(g_{\mu\nu}) \), denotes the determinant of the metric tensor \( g_{\mu\nu} \) and \( \mathcal{L} \) represents the ModMax Lagrangian \cite{kosyakov2020nonlinear}. We assume that the ModMax Lagrangian in three-dimensional spacetime takes a form analogous to the ModMax Lagrangian in four-dimensional spacetime, expressed in by the following relation,
\begin{equation}
    \mathcal{L} = X \cosh \gamma - \sqrt{X^2 + Y^2} \sinh \gamma,
\end{equation}
where, \( \gamma \) is a dimensionless parameter known as the ModMax parameter. In the ModMax Lagrangian, \( X \) and \( Y \), respectively, are a true scalar and a pseudoscalar. They are defined as,
\begin{equation}
    X = \frac{F}{4},
\end{equation}
\begin{equation}
    Y = \frac{Fe}{4}
\end{equation}
where \( F = F_{\mu\nu} F^{\mu\nu} \) is the Maxwell invariant and \( F_{\mu\nu} \) is called the electromagnetic tensor field. \( F_{\mu\nu} \) can be calculated as given,
\begin{equation}
    F_{\mu\nu} = \partial_{\mu} A_{\nu} - \partial_{\nu} A_{\mu},
\end{equation}
where, \( A_{\mu} \) is the gauge potential. In addition, \( Fe = F_{\mu\nu} Fe^{\mu\nu} \), and \( Fe_{\mu\nu} = \frac{1}{2} \epsilon_{\rho\lambda\mu\nu} F^{\rho\lambda} \). It is notable that the ModMax Lagrangian reduces to the linear Maxwell theory, i.e., \( \mathcal{L} = \frac{F}{4} \), when \( \gamma = 0 \).
The BTZ BH solution within framework of the ModMax theory is therefore given by \cite{eslam2024thermodynamics},
\begin{equation} \label{metric}
ds^2 = -\psi(r) \, dt^2 + \frac{dr^2}{\psi(r)} + r^2 \, d\phi^2,
\end{equation}
where the metric function is defines as,
\begin{equation} \label{mfunction}
\psi(r) = -m_0 - \Lambda r^2 - 2q^2 e^{-\gamma} \ln \left( \frac{r}{r_0} \right).
\end{equation}
Here, $m_{0}$ is an integral constant which is related to the total mass of BH. $r_{0}$ is an arbitrary length parameter and $q$ represents the charge of BH. $\Lambda$ is the cosmological constant already defined earlier. Throughout the manuscript, we have considered its negative value. $\gamma$ is a ModMax parameter and in absence of this parameter the metric turns into BTZ BH in Einstein-$\Lambda$-Maxwell theory.
\subsection{Horizon Structure}
\label{sec2.1}
We analyze the behavior of the metric function as a function of radial distance to study the nature of the horizons. By fixing the charge and the cosmological constant parameters, we observe that for larger values of the ModMax parameter, two horizons are present: an inner horizon and an outer (BH) horizon. Additionally, when the ModMax parameter is held constant and the cosmological constant is varied, we find that decreasing the value of the cosmological constant results in the existence of only a single horizon. However, two horizons are observed within a specific range of parameter values. As a result, small charged AdS BTZ BHs with large values of the ModMax parameter and mass have two roots, which are inner root and event horizon, respectively.
\begin{figure}[H]
	\centering 
	\includegraphics[width=0.45\textwidth, angle=0]{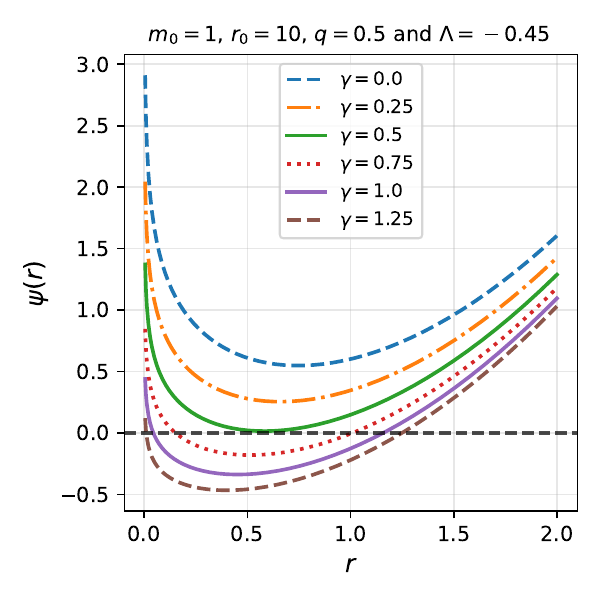}
        \includegraphics[width=0.45\textwidth, angle=0]{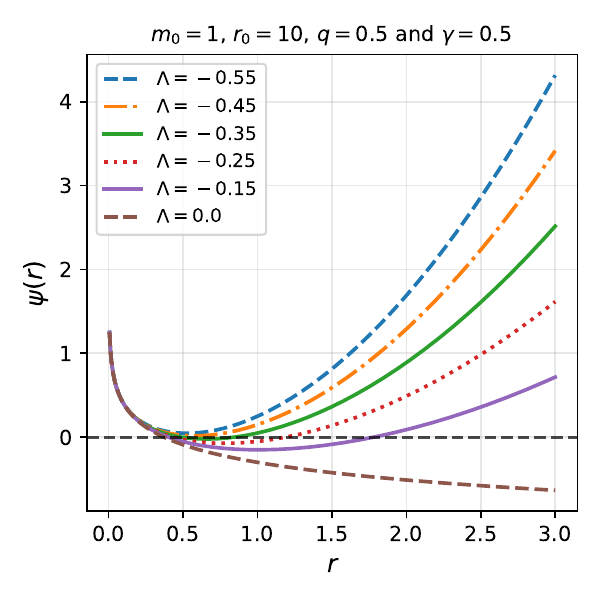}
	\caption{The variation of metric function with radial distance for different values of ModMax and Cosomological constant parameter.} 
	\label{horizon}%
\end{figure}
The analytical solution of the BH horizon can be obtained using the method outlined in \cite{Hendi:2010px}. Accordingly, the horizons radii of the BH are determined as follows,
\begin{equation} \label{hrz1}
r_+ = r_0 \exp 
\left[
-\frac{1}{2} \ W_{-1}\left(
\frac{\Lambda}{q^2 \ e^{-\gamma}}r_0^2 \  e^{\left(-\frac{m_0}{q^2 \ e^{-\gamma}}\right)}
\right) - \frac{m_0}{2 \ q^2 \ e^{-\gamma}}
\right],
\end{equation}
\begin{equation} \label{hrz2}
r_{-} = r_0 \exp 
\left[
-\frac{1}{2} \ W_{0}\left(
\frac{\Lambda}{q^2 \ e^{-\gamma}}r_0^2 \  e^{\left(-\frac{m_0}{q^2 \ e^{-\gamma}}\right)}
\right) - \frac{m_0}{2 \ q^2 \ e^{-\gamma}}
\right],
\end{equation}
where \( W_{0} \) and \( W_{-1} \) denote the two real branches of the Lambert \( W \) function. By convention, these are defined as  
\[
W_{0} : \left[ -\frac{1}{e}, \infty \right[ \to \left[ -1, \infty \right[,
\quad
W_{-1} : \left[ -\frac{1}{e}, 0 \right[ \to \, ] -\infty, -1[,
\]
with the convention that  
\[
W_{0}\!\left(-\frac{1}{e}\right) = W_{-1}\!\left(-\frac{1}{e}\right) = -1.
\]
 For a comprehensive and detailed discussion on the Lambert \(W\) function, readers are referred to \cite{corless1996lambert}.  $r_{+}$ and $r_{-}$ represent the outer and inner horizons of BH respectively.
The existence of real and distinct BH horizons, as given by Eqs.~(\ref{hrz1}) and (\ref{hrz2}), is ensured when the following condition holds:
\begin{equation}
-\frac{1}{e} < \frac{\Lambda}{q^{2} e^{-\gamma} } r_{0}^{2} \exp\!\left(-\frac{m_{0}}{q^{2} e^{-\gamma}}\right) < 0.
\end{equation}
From the above analysis, it is evident that the BH horizon is influenced by various parameters, such as $\gamma$, $q$, and other BH characteristics. The effect of these parameters on the horizon is clearly illustrated in the photon trajectories, where the BH horizon is represented by a black solid sphere.

\section{Null Geodesics}
\label{sec3}
\subsection{Geodesic equations and Effective Potential}
Geodesic equations can be obtained  using Lagrangian equation corresponding to a given
spacetime is below, 
\begin{equation} \label{lagrangian}
L = g_{\mu\nu} \frac{dx^\mu}{d\lambda} \frac{dx^\nu}{d\lambda} = -\epsilon 
= -\psi(r) \left(\frac{dt}{d\lambda}\right)^2 
+ \frac{1}{\psi(r)} \left(\frac{dr}{d\lambda}\right)^2 
+ r^2 \left(\frac{d\phi}{d\lambda}\right)^2, 
\end{equation}
where \(\epsilon\) takes the values \(1\) and \(0\) for massive and massless particles, respectively, and \(\lambda\) is the affine parameter for massless particles and the proper time for massive particles. The BTZ like metric admits two Killing vectors~\cite{Banados:1992gq}, namely \( \partial / \partial t \) and \( \partial / \partial \phi \). These symmetries give rise to two conserved quantities along the geodesics, which are expressed as
\begin{equation} \label{consofmotion}
E = g_{tt} \frac{dt}{d\lambda} = -\psi(r) \frac{dt}{d\lambda},
\quad
L = g_{\phi\phi} \frac{d\phi}{d\lambda} = r^2 \frac{d\phi}{d\lambda}. 
\end{equation}
From this, we then find the following geodesic equations,
\begin{equation} \label{rdot}
\left(\frac{dr}{d\lambda}\right)^2 = E^2 - \psi(r) 
\left(\frac{L^2}{r^2} + \epsilon\right),
\end{equation}
The effective potential corresponding to null geodesics then given by,
\begin{equation} \label{effp}
V_{\text{eff}} (r) = \left\{ -m_0 - \Lambda r^2 - 2q^2 e^{-\gamma} \ln \left( \frac{r}{r_0} \right)\right\} \frac{L^2}{r^2}.
\end{equation}
\begin{figure}[H]
	\centering 
	\includegraphics[width=0.45\textwidth, angle=0]{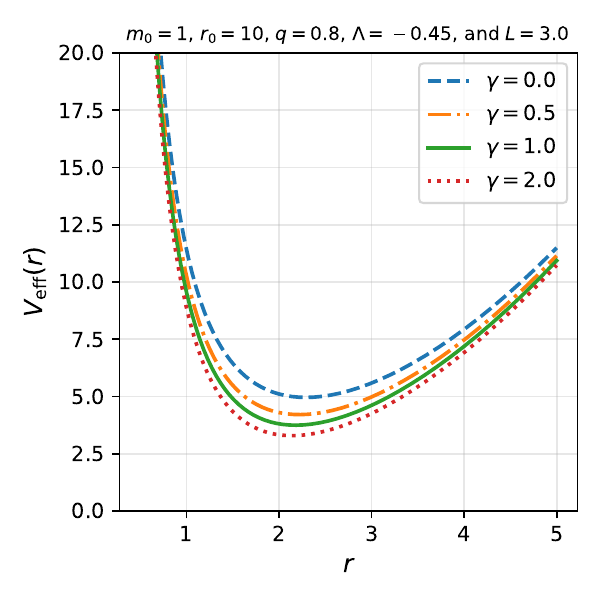}
        \includegraphics[width=0.45\textwidth, angle=0]{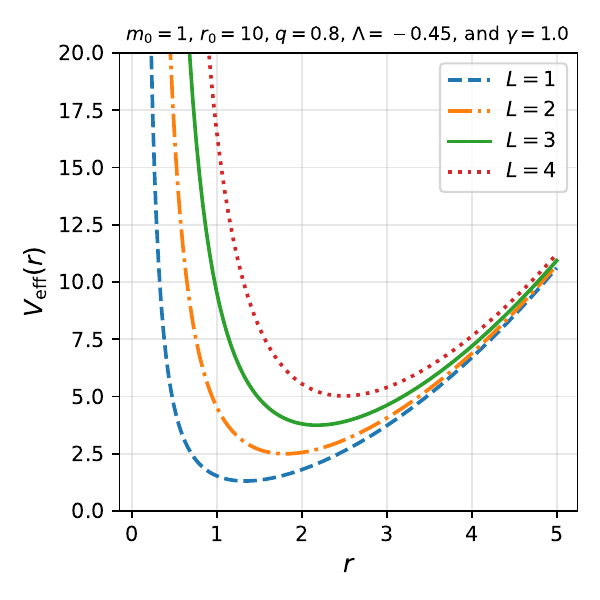}
	\caption{The variation of effective potential with radial distance for different values of ModMax parameter and and angular momentum of massless particles.} 
	\label{effective}%
\end{figure}
\noindent The variation of the effective potential with the radial distance for different ModMax parameters and angular momentum values of massless particles is shown in Fig. \ref{effective}. It is evident that the effective potential has a single minimum, suggesting the possibility of stable photon orbits around the BH. Notably, as the ModMax parameter increases, the effective potential decreases, indicating a reduction in the electrostatic repulsion between the BH and photons. This weakening of the repulsive interaction allows more particles to escape from the BH's gravitational field. Furthermore, the effective potential reaches its maximum for the highest values of angular momentum, as the increased centrifugal force pushes the particles outward, preventing them from falling into the BH. 
\subsection{Photon orbit and possible types of trajectories}
\begin{figure}[H]
	\centering 
	\includegraphics[width=0.32\textwidth, angle=0]{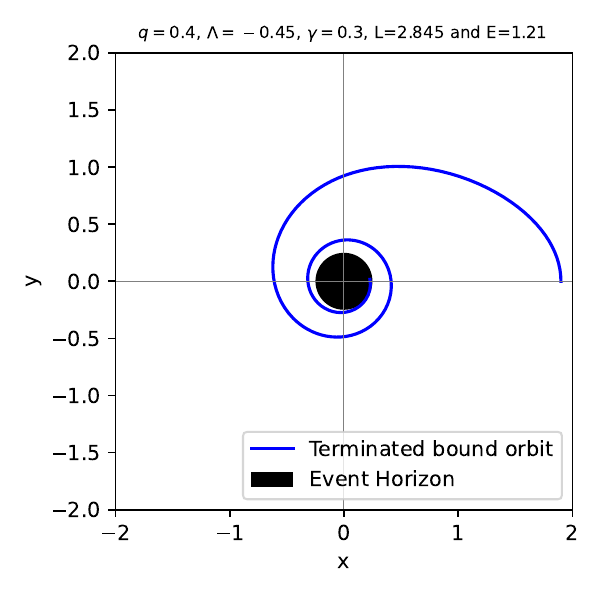}
        \includegraphics[width=0.32\textwidth, angle=0]{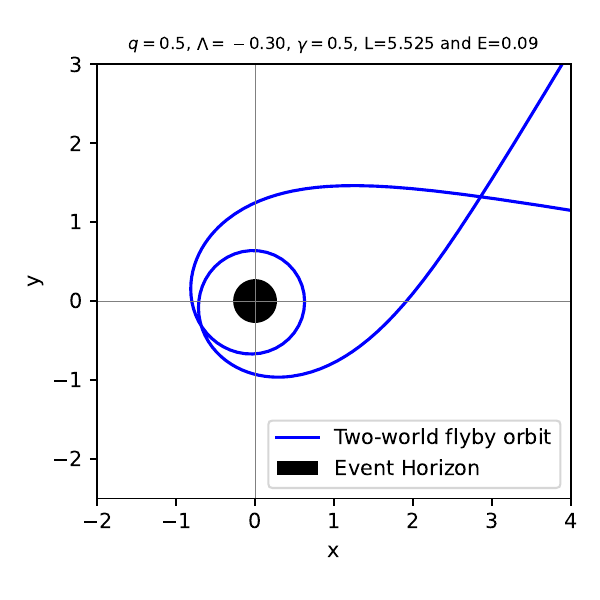}
        \includegraphics[width=0.32\textwidth, angle=0]{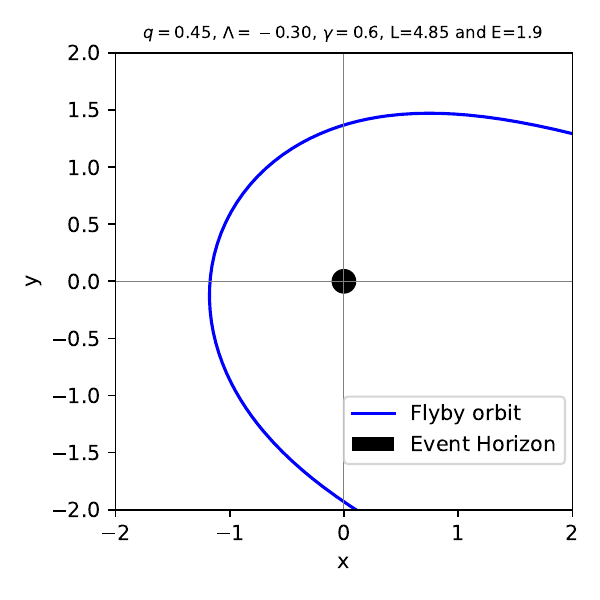}
	\caption{The figure shows three types of photon trajectories around a BH: terminated bound orbit (fall into the singularity), two-world flyby orbit (traverses near the BH and then escape to infinity) and flyby orbit (escape to infinity).} 
	\label{orbits}%
\end{figure}

The radii of the photon orbit can be determined using the condition for circular null geodesics, which requires $\dot{r} = 0 \Rightarrow V_{\text{eff}} = 0$, \quad $V'_{eff}(r) = 0$. By solving these conditions analytically, the radius of the photon orbit is obtained, and the resulting expression is given as,
\begin{equation}
    r_{ph} = r_{0} \exp\left\{ \frac{1}{2} \left( 1-\frac{m_{0}}{q^2 e^{-\gamma}} \right) \right\}.
\end{equation}
\begin{figure}[H]
	\centering 
	\includegraphics[width=0.45\textwidth, angle=0]{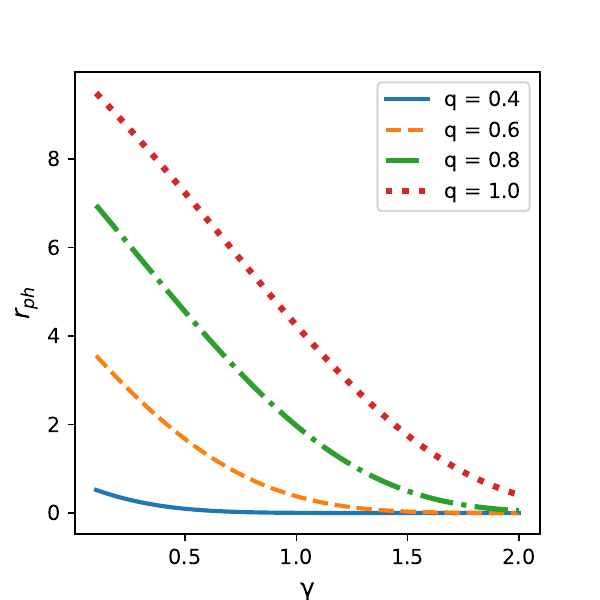}
        \includegraphics[width=0.45\textwidth, angle=0]{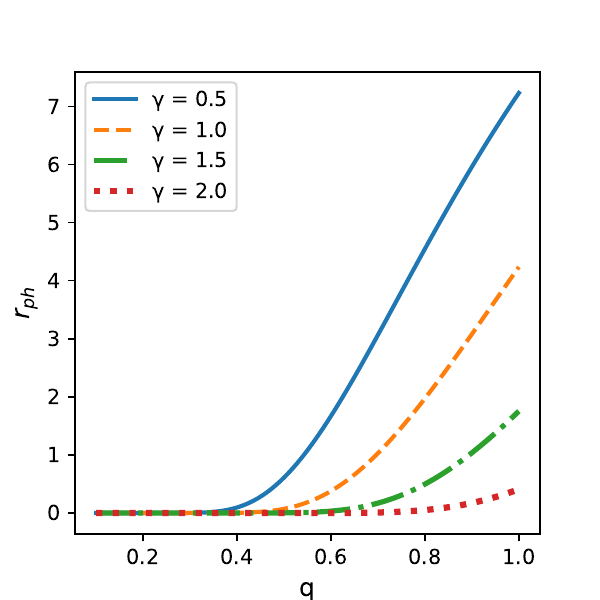}
	\caption{The graph shows the radii of photon orbit for the ModMax BTZ BH as a function of the ModMax parameter (left panel) and charge parameter (right panel). Here, we consider $m_{0}=1$ and $r_{0}=10$.} 
	\label{rphfig}%
\end{figure}
\noindent The variation in the radii of the photon orbit for different ModMax parameters and charge values is illustrated in Fig. \ref{rphfig}. It is evident that the radius of the photon orbit decreases with increasing $\gamma$, while it increases as the charge parameter grows. Additionally, we observe that for very small values of the charge parameter, the photon orbit does not exist for any value of $\gamma$. Notably, the photon orbit is independent of the cosmological constant.\\
In order to analyze the trajectories of the geodesic equation we obtain the equation of motion as,
\begin{equation}
\left(\frac{dr}{d\phi}\right)^2 = \frac{r^4}{L^2} 
\left\{E^2 - \left( -m_0 - \Lambda r^2 - 2q^2 e^{-\gamma} \ln \left( \frac{r}{r_0} \right) \right) \frac{L^2}{r^2} \right\},
\end{equation}
and, with the change of variable $u=1/r$,
\begin{equation}\label{eom}
\left(\frac{du}{d\phi}\right)^2
= \frac{1}{b^2} - u^{2}\,\psi\!\left(\frac{1}{u}\right)
= \frac{1}{b^2} 
+ m_{0}\,u^{2} + \Lambda - 2 q^{2} e^{-\gamma} \,u^{2}\ln\!\big(u r_{0}\big).
\end{equation}

where, $b=L/E$ the impact parameter. We numerically solved the Eq. \ref{eom} and depicted the possible types of trajectories in Fig. \ref{orbits}. Based on the values of angular momentum, energy, and BH parameters, two distinct types of orbit can be identified. The first type is bound orbits, where massless particles fall into the BH horizon and cannot escape. The second type is flyby orbits, where particles originating from infinity approach the BH but ultimately escape back to infinity. We identified two distinct types of flyby orbits: one is the simple flyby orbit, where the particle approaches the BH and escapes to infinity within the same asymptotic region, and the other is the two-world flyby orbit, where the particle transitions from one asymptotic region to another. The existence of two-world flyby orbits is justified by the spacetime structure of the BH, particularly in cases where the geometry includes multiple asymptotic regions, such as in extended or maximally extended spacetimes, allowing particles to traverse between these regions without being captured by the BH. A detailed analysis of the definitions and characteristics of these orbits can be found in \cite{Hackmann:2010tqa}. 
\subsection{Stability of Null Geodesics}
\label{sec3.1}
\begin{figure}[H]
	\centering 
	\includegraphics[width=1\textwidth, angle=0]{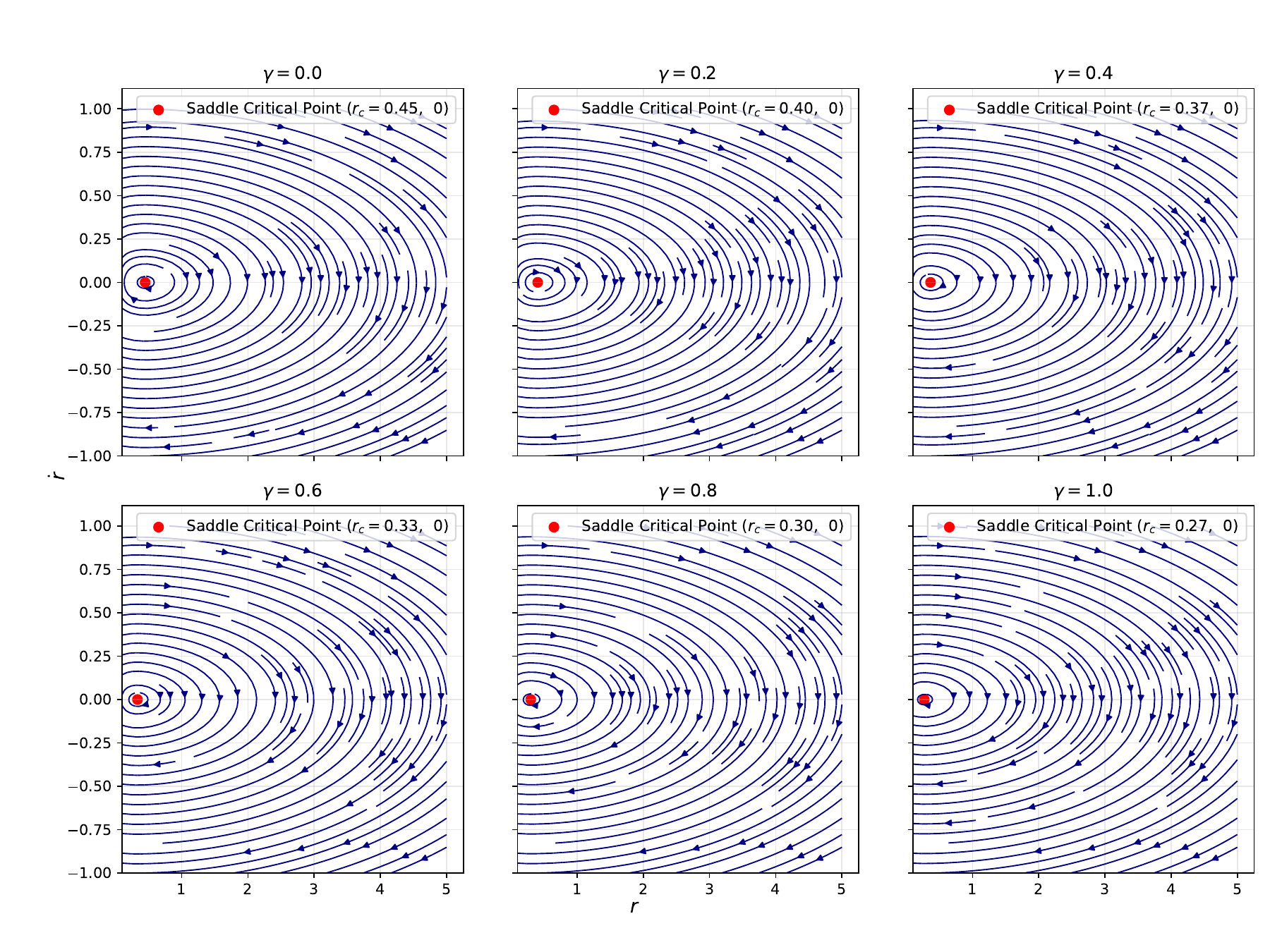}
	\caption{The graph shows the phase portrait $r$ vs $\dot{r}$ of the null geodesics for different values of the parameter $\gamma$. Here we fix the value of $m_{0}=1$,  $q=0.3$, $r_{0}=10$ and $\Lambda=-0.45$.} 
	\label{stability}%
\end{figure}

In this subsection, we analyze the stability of null circular geodesics, we construct a dynamical system and analyze its phase space structure in the $(r, \dot{r})$ plane \cite{goldhirsch1987stability}. The study of the behavior of small perturbations around the circular geodesics enables us to determine whether the orbits are stable or unstable under radial perturbations. The condition $\dot{r} = 0$ in null geodesics, implies that the radial velocity vanishes, leading to a reduction of the phase space dynamics to a two-dimensional subspace characterized by the variables $r$ and its conjugate momentum. Analyzing the phase flow dynamics in the $(r, \dot{r})$ plane allows us to identify the critical point associated with the photon orbit radius $(r_{c},0)$. To advance this analysis, we differentiate Eq. \ref{rdot} with $\epsilon=0$, eliminating $\dot{r}$ to derive the following expression,
\begin{equation}
    \ddot{r} = -\frac{dV_{\text{eff}}}{dr}.
\end{equation}
Now we consider the coordinates $x_1 = \dot{r}$ and $x_2 = \dot{x_1}$, leading to the differential equations corresponding to these coordinates, which are given by,
\begin{equation} \label{JacobianM}
    \begin{aligned}
x_1 &= \dot{r}, \\
x_2 &= -\frac{dV_{\text{eff}}}{dr}.
\end{aligned}
\end{equation}
The Jacobian matrix $J$ corresponding to the sets of differential  Eq. \ref{JacobianM} is given by \cite{Yang:2023hci},
\begin{equation}
    J = \begin{pmatrix}
0 & 1 \\
-V_{\text{eff}}''(r) & 0
\end{pmatrix}, 
\end{equation}
where, $V_{\text{eff}}''(r)$ denotes the second derivative of the effective potential with respect to $r$. The secular equation $|J - \lambda I| = 0$ yields the eigenvalue squared as,
\begin{equation}
    \lambda_{L}^2 = -V_{\text{eff}}''(r). 
\end{equation}
Here, $\lambda_{L}$ is known as the Lyapunov exponent. It measure the average rate at which nearby trajectories converge or diverge in the phase space. $\lambda_{L}^2 > 0$ indicates a divergence between nearby trajectories, i.e., the behavior of the system exhibits both stable and unstable directions (saddle critical points). Conversely, the condition $\lambda_{L}^2 < 0$ indicates a convergence between nearby trajectories, i.e., critical point represents a stable center.\\
The phase portraits in Fig. \ref{stability} illustrate the dynamical behavior of null geodesics in the $(r, \dot{r})$ phase space for different values of the parameter $\gamma$. Each trajectory in the phase plane represents the evolution of a photon orbit under small radial perturbations, where the critical point $(r_{c}, 0)$ corresponds to the photon-sphere radius. As seen in the plots, the trajectories near the critical point exhibit the characteristic structure of a saddle, confirming the unstable nature of the circular photon orbits. With increasing $\gamma$, the saddle point shifts toward smaller radial distances, indicating that the photon orbit moves inward as the parameter $\gamma$ increases. Specifically, for $\gamma = 0$, the critical point occurs at approximately $(r, \dot{r}) = (0.45, 0)$, while for $\gamma = 1$, it shifts to $(r, \dot{r}) = (0.27, 0)$. This inward shift reflects a stronger gravitational attraction or effective curvature induced by higher $\gamma$ values, causing photons to be trapped closer to the BH.
Thus, the phase portraits provide a clear visualization of how the stability and spatial location of photon orbits are influenced by the parameter $\gamma$, demonstrating the dynamical instability of the photon orbit in the modified BH spacetime.

\subsection{Gaussian Optical Curvature}
\label{sec3.2}
In this subsection we now focus on null geodesics deflected by this BH to analyze the gaussian optical curvature using the optical metric. It is well-established that light follows null geodesics, which are characterized by the condition (i.e., \(ds^2 = 0\)). This null condition is fundamental in defining the optical metric, which effectively captures the geometry traced by light as it propagates through a curved spacetime. In this context, the optical metric provides an approximation of the Riemannian geometry experienced by photons, facilitating the study of phenomena such as GL and light deflection. Incorporating the condition of null geodesics, the associated optical metric is given by \cite{Ishihara:2016vdc},
\begin{equation} \label{opticalmetric}
dt^2 = \bar{g}_{ij} dx^i dx^j = d\tilde{r}^2 + \psi(\tilde{r})^2 d\phi^2,
\end{equation}
where
\begin{equation} \label{eq12}
    d\tilde{r} = \frac{dr}{-m_0 - \Lambda r^2 - 2q^2 e^{-\gamma} \ln \left( \frac{r}{r_0} \right)},
\end{equation}

\begin{equation} \label{eq13}
    \psi(\tilde{r}) = \frac{r}{\sqrt{-m_0 - \Lambda r^2 - 2q^2 e^{-\gamma} \ln \left( \frac{r}{r_0} \right)}}. 
\end{equation}
The equatorial plane within the optical metric represents a surface of revolution. The Christoffel symbols corresponding to the metric in Equation \ref{opticalmetric}, which are non-zero, are determined as follows,
\begin{equation}
    \tilde{\Gamma}^r_{\phi\phi} = \frac{r(r \psi'(\tilde{r}) - 2\psi(\tilde{r}))}{2}, 
\end{equation}

\begin{equation}
    \tilde{\Gamma}^\phi_{r\phi} = \frac{2\psi(\tilde{r}) - r \psi'(\tilde{r})}{2}, 
\end{equation}

\begin{equation}
    \tilde{\Gamma}^r_{rr} = -\frac{\psi'(\tilde{r})}{\psi(\tilde{r})}. 
\end{equation}
Here, the prime indicates differentiation with respect to $r$. Using the Christoffel symbols derived above, the only non-zero components of the Riemann tensor that contribute to the optical curvature are as follows,
\begin{equation}
    \tilde{R}^r_{\phi r \phi} = -\kappa \psi^2(\tilde{r}),
\end{equation}
where \(\kappa\) represents the geodesic curvature. Consequently, the Gaussian optical curvature \(\mathcal{K}\) can be expressed in terms of the Ricci scalar as
\begin{equation}
    \mathcal{K} = \frac{R}{2} = -\frac{1}{\psi(\tilde{r})} 
\left[
\frac{dr}{d\tilde{r}} \frac{d}{dr} 
\left(\frac{dr}{d\tilde{r}}\right) \frac{d\psi}{dr} 
+ \frac{d^2 \psi}{dr^2} \left(\frac{dr}{d\tilde{r}}\right)^2
\right].
\end{equation} 
Corresponding to Eqs. \ref{eq12} and \ref{eq13}, the Gaussian optical curvature can ultimately be expressed in the following precise form,
\begin{equation}\label{gaucurv}
\mathcal{K}(r) \approx 
\Lambda m_{0}
+ 2 \Lambda q^{2} e^{-\gamma} \ln\!\left( \frac{r}{r_{0}} \right)
- 3 \Lambda q^{2} e^{-\gamma}
- \frac{m_{0} q^{2} e^{-\gamma}}{r^{2}}
+ \mathcal{O}\!\left(q^{4},\, m_{0}^{2}\right).
\end{equation}
The obtained expression for Gaussian curvature indicates that it is influenced by various parameters, including mass, charge, ModMax, cosmological constant, and length parameter.

\section{Deflection of Light}
\label{sec4}
\begin{figure}[H]
	\centering 
	\includegraphics[width=0.45\textwidth, angle=0]{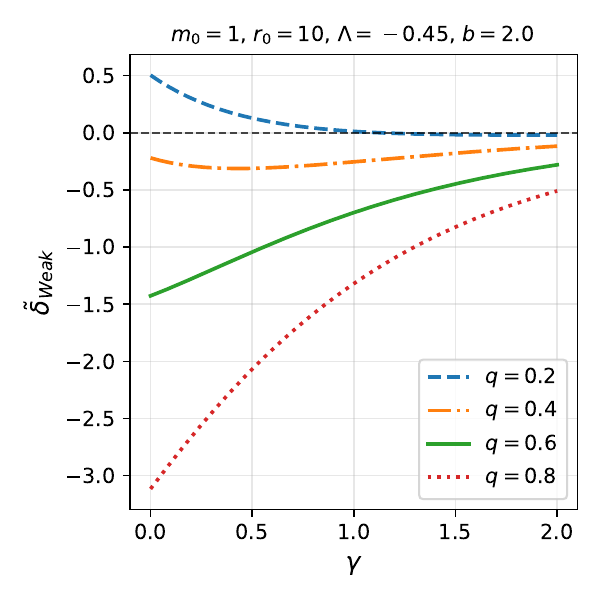}
        \includegraphics[width=0.45\textwidth, angle=0]{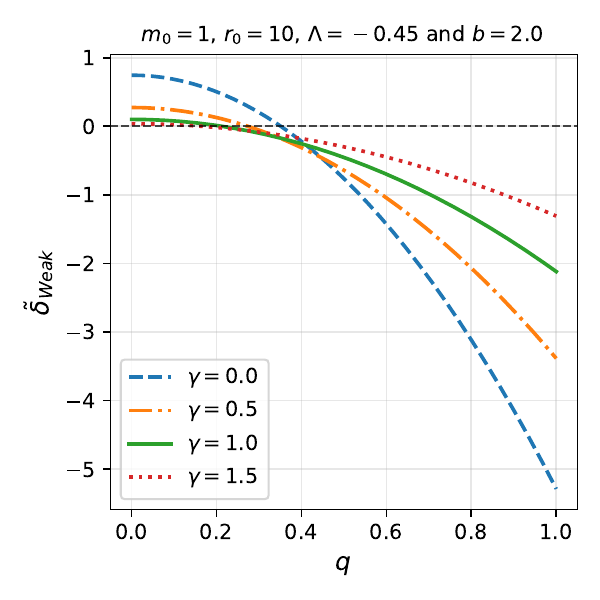}
	\caption{The variation of deflection angle with ModMax parameter (left panel) and charge parameter (right panel) including the different values of these parameters.} 
	\label{dangle2}%
\end{figure}

\begin{figure}[H]
	\centering 
	\includegraphics[width=0.45\textwidth, angle=0]{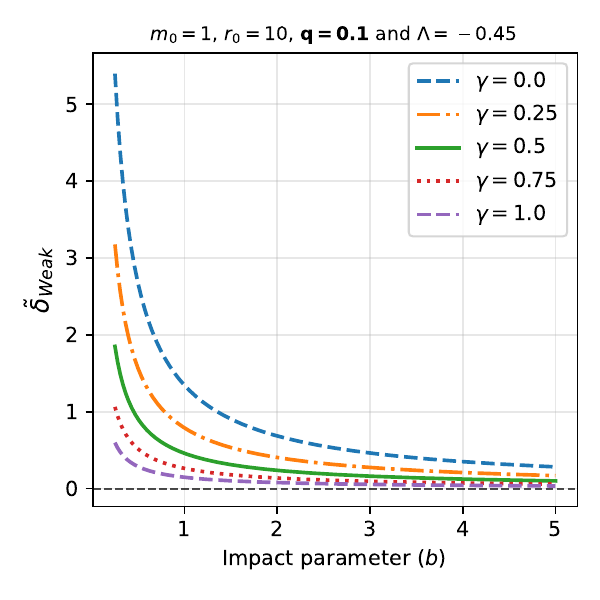}
        \includegraphics[width=0.45\textwidth, angle=0]{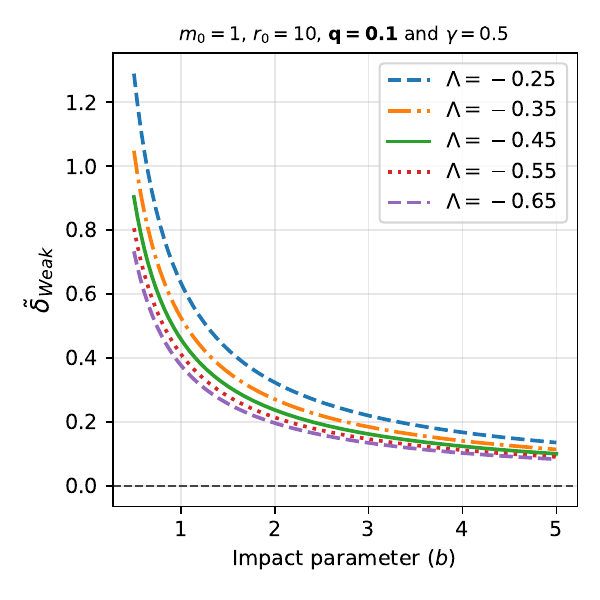}
	\caption{The variation of deflection angle with impact parameter for different values of $\gamma$ (left panel) and $\Lambda$ (right panel) for low value of charge parameter. } 
	\label{dangle3}%
\end{figure}

\begin{figure}[H]
	\centering 
	\includegraphics[width=0.45\textwidth, angle=0]{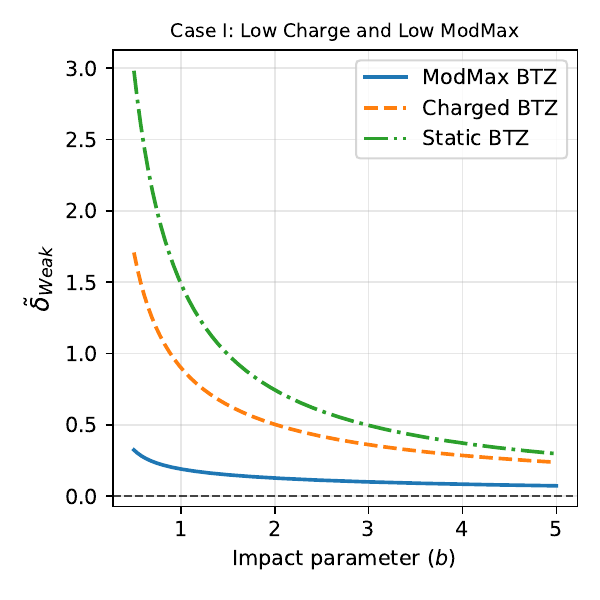}
        \includegraphics[width=0.45\textwidth, angle=0]{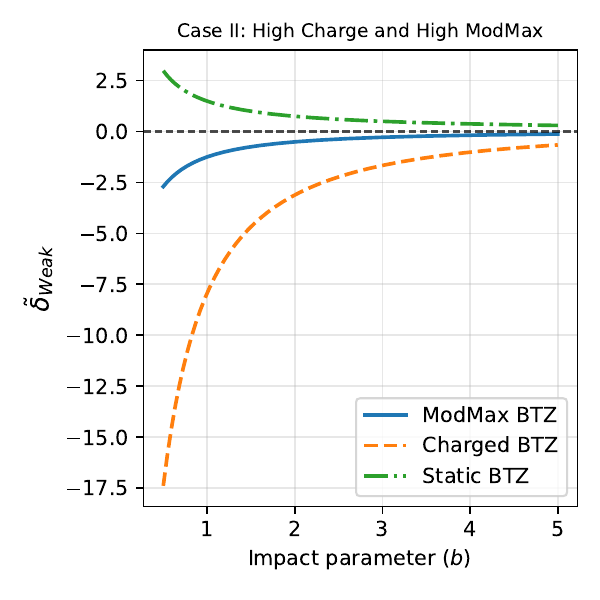}
	\caption{The variation of deflection angle with impact parameter and comparison of other BTZ BH solution. For Case I we consider $\gamma=0.5$ and $q=0.2$ and for Case II, $\gamma=1.5$ and $q=0.8$.  } 
	\label{dangle4}%
\end{figure}

In this section, we compute the deflection angle of a ModMax BTZ BH using the Gauss-Bonnet theorem (GBT). The GBT provides a connection between the intrinsic differential geometry of a metric associated with its topology within a regular domain is expressed as \cite{Gibbons:2008rj},  
\begin{equation}
    \iint_{U_R} \mathcal{K} \, dS + \oint_{\partial U_R} \kappa \, ds + \sum_{z} \beta_z = 2\pi \chi(U_R),
\end{equation}
where \( U_R \subset \Sigma \) is a regular domain of a two-dimensional surface. \(\kappa\) denotes the geodesic curvature, defined as \(\kappa = \bar{g}(\nabla_{\dot{\sigma}} \dot{\sigma}, \ddot{\sigma})\), where \(\sigma\) is a smooth curve of unit speed such that \(\bar{g}(\dot{\sigma}, \dot{\sigma}) = 1\), and \(\ddot{\sigma}\) is the unit acceleration vector. \(\beta_z\) represents the exterior angle at the \(z\)-th vertex, and \(\chi\) is the Euler characteristic number.  
In the respective limit \( R \to \infty \) (for the curve \( C_R \)), the geodesic curvature simplifies to \(\kappa(C_R) = |\nabla_{\dot{C}_R} \dot{C}_R|\). The radial component of the geodesic curvature can be expressed as,  
\begin{equation}
    \left(\nabla_{\dot{C}_R} \dot{C}_R\right)^r = \dot{C}_R^\phi \partial_\phi \dot{C}_R^r + \Gamma^r_{\phi\phi} \left(\dot{C}_R^\theta\right)^2.
\end{equation}
For sufficiently large \(R\), the curve \(C_R\) is defined by \(r(\phi) = R = \text{constant}\), leads to,
\begin{equation}
    \left(\dot{C}_R^\phi\right)^2 = \frac{1}{\Psi^2(\tilde{r})}.
\end{equation}
In the context of optical geometry, the geodesic curvature can be expressed using the Christoffel symbols as,
\begin{equation}
    \left(\nabla_{\dot{C}_R} \dot{C}_R\right)^r \to \frac{1}{R}. 
\end{equation}
This implies that \(\kappa(C_R) \to \frac{1}{R}\). Using the optical metric defined in \ref{opticalmetric}, we find \(dt = R d\phi\). Consequently,
\begin{equation}
    \kappa(C_R) dt = \lim_{R \to \infty} [\kappa(C_R) dt] = \lim_{R \to \infty} \left[\frac{1}{2 \sqrt{\bar{g}_{rr} \bar{g}_{\phi\phi}} \left(\frac{\partial \bar{g}_{\phi\phi}}{\partial r}\right)}\right] d\phi = d\phi.
\end{equation}
Considering all these discussions, the Gauss-Bonnet theorem can be expressed as,
\begin{equation}
    \iint_{U_R}  \mathcal{K} \, dS + \oint_{\partial U_R} \kappa \, dt \Big|_{R \to \infty} =
\iint_{S_\infty}  \mathcal{K} \, dS + \int_{\pi + \tilde{\delta}}^{0} d\phi. 
\end{equation}
In the weak deflection regime, the light ray at zeroth order follows a straight trajectory given by \(r(\phi) = \frac{b}{\sin \phi}\), where $b$ denotes the impact parameter. From this, the deflection angle can be derived as,
\begin{multline}
    \tilde{\delta} = -\int_0^\pi \int_{b / \sin \phi}^\infty  \mathcal{K} \, dS 
    = -\int_0^\pi \int_{b / \sin \phi}^\infty  \mathcal{K} \sqrt{\det \bar{g}} \, d\tilde{r} d\phi \\
    = -\int_0^\pi \int_{b / \sin \phi}^\infty \frac{ \mathcal{K} r}{\psi(r)^{3/2}} \, d\tilde{r} d\phi.
\end{multline}
Therefore, the deflection angle can be express as follows,
\begin{equation}\label{bendingangle}
\begin{aligned}
\tilde{\delta}
&= -\frac{2 m_{0}}{b\sqrt{-\Lambda}}
-\frac{2q^{2}e^{-\gamma}}{b\sqrt{-\Lambda}}
\Big(2\ln b - 2\ln(2r_{0}) + 1\Big) 
+\frac{4}{9}\,\frac{m_0 q^{2} e^{-\gamma}}{b^{3}(-\Lambda)^{3/2}}
+\mathcal{O}\!\left(m_{0}^{2},\,q^{4},\,m_0^2 q^2\right).
\end{aligned}
\end{equation}
The deflection angle obtained in Eq. \ref{bendingangle} demonstrates that it is directly proportional to the ModMax parameter and inversely proportional to the cosmological constant. When the ModMax parameter is set to zero, the expression of the deflection angle reduces to that of the charged BTZ BH, aligning with the findings of Upadhyay et al. \cite{Upadhyay:2023yhk}.
\begin{figure}[H]
	\centering 
	\includegraphics[width=0.45\textwidth, angle=0]{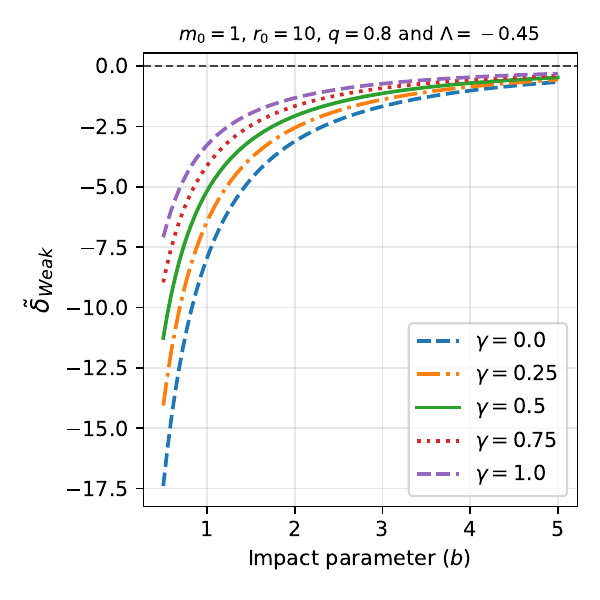}
        \includegraphics[width=0.45\textwidth, angle=0]{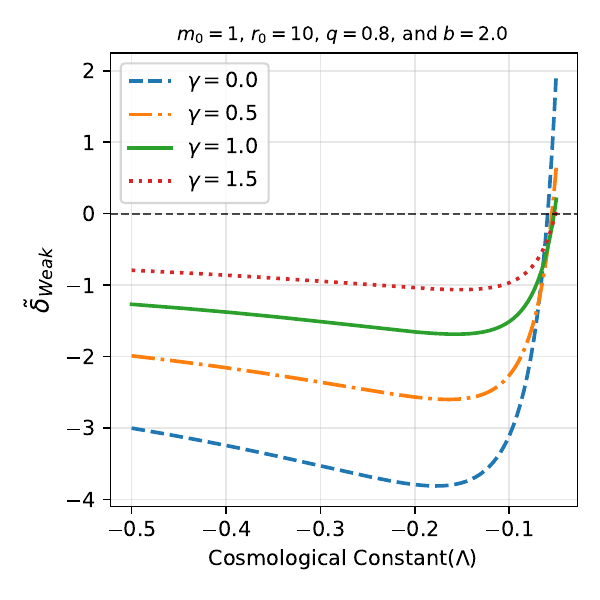}
	\caption{The variation of deflection angle with impact parameter (left panel) and cosmological constant (right panel) for different values of $\gamma$.} 
	\label{dangle1}%
\end{figure}
\noindent We analyze the behavior of the deflection angle as a function of the impact parameter and other BH parameters in detail. The graphical representations of these analyses are shown in Figs.~\ref{dangle1} to \ref{dangle4}. For large values of the charge parameter, the deflection angle increases with the impact parameter, highlighting the interplay between gravitational attraction and the repulsive effects introduced by the charge. Conversely, for decreasing values of the cosmological constant $\Lambda$, the deflection angle decreases, indicating a reduction in the curvature of spacetime in lower-$\Lambda$ environments. When examining the influence of the ModMax parameter, we observe a nuanced behavior. For the lowest values of the charge parameter, the deflection angle decreases with increasing ModMax parameter. In contrast, for higher values of the charge parameter, the deflection angle increases. This suggests that the combined effects of charge and ModMax parameter are interdependent, with the charge playing a dominant role in determining the net deflection behavior.
Interestingly, beyond a specific critical value of the charge parameter, the deflection angle shifts to the negative region. This shift can be attributed to the additional repulsive force arising from the strong electromagnetic contribution in highly charged BHs. To explore this phenomenon further, we analyzed the deflection angle for lower charge parameter values. For these cases, the deflection angle decreases with the impact parameter for varying $\gamma$ and $\Lambda$. Notably, higher values of $\gamma$ correspond to lower deflection angles, indicating a reduction in the gravitational pull of the BH due to the increasing influence of the ModMax parameter. Additionally, larger values of the cosmological constant $\Lambda$ result in greater deflection, consistent with the enhanced spacetime curvature. Finally, we compare our results with other BH solutions, specifically static BTZ BHs as shown in Fig. \ref{dangle4}, under two scenarios as follows; \textbf{CaseI: Low charge and low ModMax parameter:} In this case, the static BTZ BH exhibits the largest deflection angle, while the ModMax BTZ BH produces the smallest deflection. \textbf{Case II: High charge and high ModMax parameter:} For this configuration, both charged and ModMax BTZ BHs exhibit negative deflection angles, reflecting the dominance of repulsive effects. However, the ModMax BTZ BH displays a larger deflection than the charged BTZ BH, signifying a more complex interaction between the ModMax parameter and the charge in determining the photon trajectories.

\section{Summary and conclusions}
\label{sec5}
In this work, we have investigated the propagation of massless particles in the background of the ModMax BTZ BH, analyzing its horizon structure, null geodesics, photon motion, stability properties, and the corresponding deflection angle using the Gauss--Bonnet theorem. Our analysis of the metric function revealed that small charged AdS BTZ BHs with large ModMax parameter values exhibit two distinct horizons, independent of the cosmological constant, thus enriching the classical picture of horizon structure in lower-dimensional gravity. The study of null geodesics, based on the Lagrangian approach, indicated that unstable photon orbits are not possible, although marginally stable circular orbits may exist. We observed that the photon orbit radius decreases with increasing ModMax parameter $\gamma$ while increasing with the charge parameter $q$. Depending on the energy, angular momentum, and BH parameters, photon trajectories can correspond to bound or flyby orbits, reflecting a rich interplay between geometry and particle motion.  

The stability of null geodesics, examined using a dynamical systems approach, revealed that both $\gamma$ and $q$ strongly influence orbital behavior, leading to new dynamical features compared to standard BTZ spacetimes. The deflection angle was found to increase with the impact parameter for large charges but to decrease with a smaller cosmological constant, while higher values of the ModMax parameter reduced the gravitational pull and hence the deflection. Interestingly, for low charge values, the deflection angle decreases with increasing ModMax parameter, whereas for higher charges it increases. Beyond a critical charge, the deflection angle becomes negative, signifying the dominance of repulsive electromagnetic effects. A comparative analysis with static and charged BTZ BHs showed that, while the static case yields larger deflection for low charges and small ModMax values, at higher charges both charged and ModMax BTZ spacetimes exhibit negative deflection, with the ModMax case presenting a stronger repulsive character.  

These findings carry observational significance in an indirect but important sense. Since $(2+1)$-dimensional BHs are not astrophysical objects, direct confrontation with observational data, such as those from the Event Horizon Telescope (EHT), is not possible. Nevertheless, the trends observed in our analysis provide valuable analogies: they highlight how nonlinear electrodynamics modifications can manifest in photon trajectories and lensing effects. Such lower-dimensional models thus serve as theoretical laboratories that help us understand how modifications from nonlinear dynamics may alter light propagation when compared to the standard Schwarzschild or BTZ cases, offering phenomenological insights that can guide the interpretation of astrophysical BH observations in realistic spacetimes.

The motivation for considering the ModMax electrodynamics framework lies in its special theoretical properties: it represents the unique one-parameter, conformally invariant extension of Maxwell’s theory that preserves duality symmetry while incorporating nonlinear effects. Unlike other nonlinear electrodynamics models applied to BTZ BHs, the ModMax theory ensures consistency with conformal symmetry, which is a cornerstone in both field theory and lower-dimensional gravity. Studying the BTZ--ModMax system therefore provides a well-motivated and novel arena to explore how such nonlinear, symmetry-preserving modifications influence BH physics. Our work fills an important gap by connecting the structural features of ModMax electrodynamics with the dynamics of photon propagation in lower-dimensional BHs.

In summary, this study not only enriches the theoretical landscape of nonlinear electrodynamics in (2+1)-dimensional gravity but also opens a pathway for using photon propagation and lensing phenomena as potential probes of novel modifications to electrodynamics.

\section*{Acknowledgments}
The author would like to thank the anonymous referees for their valuable comments. Author sincerely acknowledges IMSc for providing excellent facilities and a conducive environment to carry out his research as a postdoctoral fellow at the Institute. The author is very thankful to Prof. Hemwati Nandan from the Department of Physics, Hemwati Nandan Bahuguna Garhwal University, Srinagar, Uttarakhand, India, for his kind support and help during this research.

\bibliographystyle{unsrt} 
\bibliography{modmax}






\end{document}